# A THRESHOLD SECURE DATA SHARING SCHEME FOR FEDERATED CLOUDS

K.Venkataramana[1], Dr.M.Padmavathamma[2]

[1]Research Scholar, Department of Computer Science, S.V.University, Tirupati, A.P, India
Email: ramanakv4@gmail.com

[2]Research Supervisor & Head, Department of Computer Science, S.V.University, Tirupati, A.P, India
Email: prof.padma@yahoo.com

**Abstract:** Cloud computing allows users to view computing in a new direction, as it uses the existing technologies to provide better IT services at low-cost. To offer high QOS to customers according SLA, cloud services broker or cloud service provider uses individual cloud providers that work collaboratively to form a federation of clouds. It is required in applications like Real-time online interactive applications, weather research and forecasting etc., in which the data and applications are complex and distributed. In these applications secret data should be shared, so secure data sharing mechanism is required in Federated clouds to reduce the risk of data intrusion, the loss of service availability and to ensure data integrity. So In this paper we have proposed zero knowledge data sharing scheme where Trusted Cloud Authority (TCA) will control federated clouds for data sharing where the secret to be exchanged for computation is encrypted and retrieved by individual cloud at the end. Our scheme is based on the difficulty of solving the Discrete Logarithm problem (DLOG) in a finite abelian group of large prime order which is NP-Hard. So our proposed scheme provides data integrity in transit, data availability when one of host providers are not available during the computation.

## I. INTRODUCTION

Cloud computing can be viewed as a new paradigm for dynamic and controlled provisioning of sharable computing resources, maintained by state-of-the-art data centers based on network of Virtual Machines running on high powered physical machines. NIST[1] defines Cloud computing whose main design aim is to provide convenient, on-demand, network access to a shared pool of configurable computing resources (e.g. networks, servers, storage, applications, and services), which can be rapidly provisioned and released with minimal management effort or service provider interactions. Cloud can be deployed in public, private or hybrid models which provides services in various forms like Software as a Service-SaaS (e.g. Google apps, 2011), Platform as a Service-PaaS (e.g. Google app engine (2011), Microsoft's Azure (Azure services platform, 2011)) and Infrastructure as Service-IaaS (e.g. Amazon web services, 2011(AWS); Eucalyptus, 2011; Open Nebula (OpenNebula, 2011).To deliver the services efficiently cloud should possess the characteristics like Resource pooling, Virtualization, Multi-tenancy, On-demand self-service, Rapid elasticity ,metered service etc., as show in Fig-1.

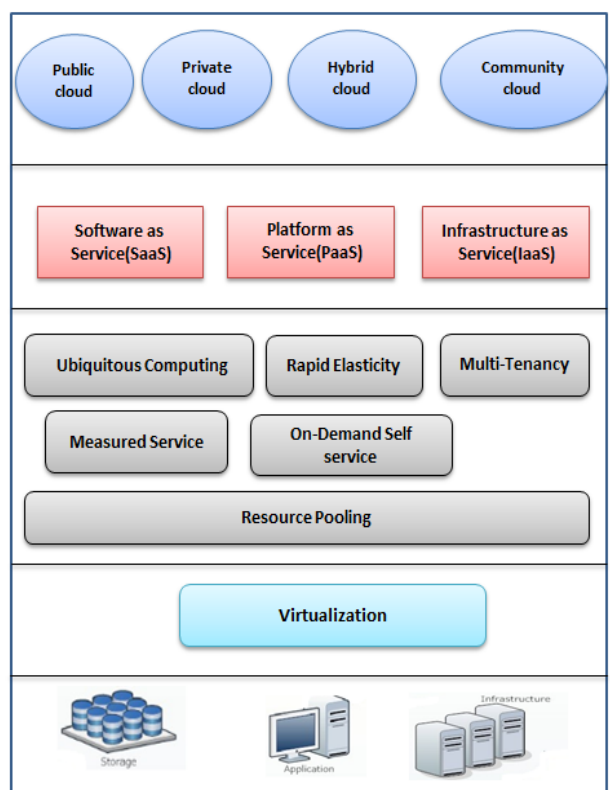

*Figure 1: Cloud Computing Model*

Slow access to data, applications, and Web pages frustrates employees and customers alike, and some performance problems and bottlenecks can even cause application crashes and data losses. So as to improve the performance, providers has to increase computing resources by their aggregated capabilities to provide infinite computing services through federation and interoperability.

As cloud computing evolves, the vision of federated clouds across which Communications, data, and services can move easily within and across several cloud infrastructures—adds another layer of complexity to security equation. Even though federated Cloud paradigm aims to provide flexible and reliable services composed of a mixture of internal and external mini-clouds, but this heterogeneous nature is also fuelling the security concerns of the customers. To allay the fears and deal with the threats associated with outsourcing data and applications to the Cloud, new methods for security assurance are urgently required. Cloud providers should address privacy and security issues as a matter of high and urgent priority. In this paper among the various security issues we consider the issue of exchanging of private data between the clouds in federation securely.

The purpose of this paper is to provide a new data sharing scheme for federated clouds which comprises various host providers which ensures privacy and availability of data. The remainder of this paper is organized as follows Section-2 summarizes previous work in the area of federated computing and its security. Section-3 introduces the federation computing, technologies and various security issues. Section- 4 specifies the proposed model and Section-5 provides working mechanism of the model. In Section-6 we have given results for the scheme and final section we have given our conclusions along with future work.

## II. RELATED WORK

As in [3] Federation is the ability of multiple independent resources to act like a single resource. Cloud computing itself is a federation of resources, so the many assets, identities, configurations and other details of a cloud computing solution must be federated to make cloud computing practical. Also many issues like trust, Identity access management, Signing-in has been discussed regarding Federation of clouds.

Buyya et al. in [4] suggests a cloud federation oriented, just-in-time, opportunistic and scalable application services provisioning environment called InterCloud. As a result Cloud application service (SaaS) providers will have difficulty in meeting QoS expectations for all their consumers. Hence, they would like to make use of services of multiple Cloud infrastructure service providers who can provide better support for their specific consumer needs. This kind of requirements often arises in enterprises with global operations and applications such as Internet service, media hosting, and Web 2.0 applications. This necessitates building mechanisms for federation of Cloud infrastructure service providers for seamless provisioning of services across different Cloud providers.

In paper by Subashini and kavitha[5], has discussed various security issues at various service models like Data security, Network security, Data locality, Data integrity, Data segregation, Data access, Authentication and authorization. Cloud computing has significant implications for the privacy of personal information as well as for the confidentiality of business and governmental information. In the case of federated clouds this becomes more serious issue that is to be addressed. For computation exchange of data between clouds in federation is necessary so both privacy and integrity of data should be considered.

Even within the cloud provider's internal network, encryption and secure communication are essential, as the information passes between countless, disparate components through network domains with unknown security, and these network domains are shared with other organizations of unknown reputability[6].The confidentiality of sensitive data must be protected from mixing with network traffic with other cloud hosts. If the data is shared between multiple users or clouds , the CSP must ensure data integrity and consistency. The CSP must also protect all of its cloud service consumers from malicious activities or data modification [7-8].

In [10] Mohammed Abdullatif et.al, has discussed about data privacy in DAAS. In their paper Shamir's secret sharing mechanism has been used for securing data , so that individual data values will not be visible to the service provider and provider can recover data in case of data loss. By above literature study we have proposed this scheme for secure data sharing in federated clouds which ensures that secret data used in computation is not visible to anyone except to owner of data ie., one of the cloud host provider who participates in computation by sharing data and avoids modification of data due to malicious host.

## III. FEDERATION COMPUTING

Cloud federation brings together different service providers and their offered services so that many Cloud variants can be tailored to match different sets of customer requirements. Cloud provider can provide resources to satisfy complex application request only if he holds infinite resources at his premises. Since this is not the case, so providers need to collaborate to be able to fulfill requests during peak demands and negotiate the use of idle resources with other peers. This is the goal of federation. The main purpose of moving to federated clouds is to improve what was offered in single clouds by distributing reliability, trust, and security among multiple cloud providers.

When increasing resources on the cloud to restore or improve application performance, administrators can scale either horizontally (out) or vertically (up), depending on the nature of the resource constraint. Vertical scaling (up) entails adding more resources to the same computing pool—for example, adding more RAM, disk, or virtual CPU to handle an increased application load. Horizontal scaling (out) requires the addition of more machines or devices to the computing platform to handle the increased demand. Scalability is the inherent feature of cloud computing which has at least two dimensions, namely horizontal cloud scalability and vertical cloud scalability [2]. Horizontal cloud scalability is the ability to connect and integrate multiple clouds to work as one logical cloud.

For instance, a cloud providing calculation services (*calculation cloud*) can access a cloud providing storage services (*storage cloud*) to keep intermediate results. Two calculation clouds can also integrate into a larger calculation cloud. Vertical cloud scalability can be used to improve the capacity of a cloud by enhancing individual existing nodes in the cloud (such as providing a server with more physical memory) or improving the bandwidth that connects two nodes.

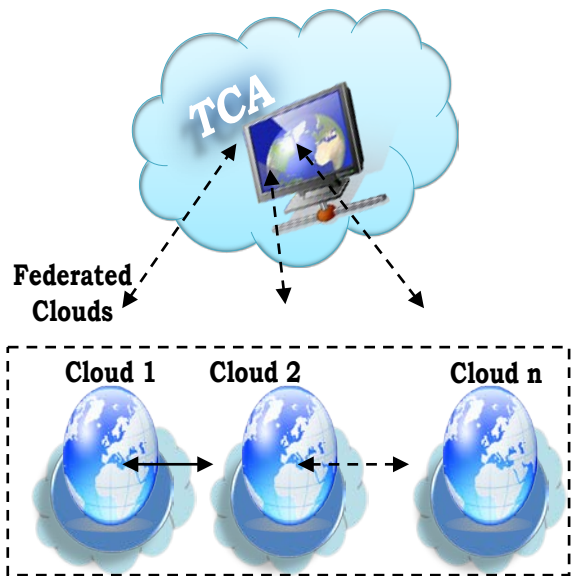

*Figure 2: Federated Clouds*

### A. Cloud Federation Technologies

As discussed in [9] the following technologies provide mechanisms which support Cloud services and even federation. Such as, Open Nebula provides an open-source and extensible architecture that can be modified to fit an individual Cloud. It can be leveraged by adding APIs and plug-ins to the existing architecture in order to facilitate inter-Cloud communication at different layers of the service stack. Eucalyptus is also an open-source framework that uses storage and computational infrastructure to provide a Cloud computing platform. Eucalyptus provides a modular, extensible framework with an Amazon EC2 compatible interface which can be utilized for federation at the IaaS layer. CometCloud is an autonomic computing engine that enables the dynamic and on-demand federation of Clouds as well as the deployment and execution of applications on these federated environments. It supports heterogeneous and dynamic Cloud infrastructures, enabling the integration of public/private Clouds and autonomic Cloud bursts, i.e., dynamic scale-out to Clouds to address dynamic workloads. Conceptually, CometCloud is composed of a programming layer, service layer, and infrastructure layer.

### B. Security issues in Federated Clouds

All the above technologies does not specify any security related measures for federated environment at any service layer, to address the data integrity, data availability and sharing. Federated clouds pose challenges like whether the client or other cloud is servicing according to SLA agreements. The diversity and flexibility of the capabilities envisioned by Inter-cloud enabled federated Cloud computing model, combined with the magnitudes and uncertainties of its components, pose difficult problems and challenges in effective provisioning and delivery of application services in an efficient and secured manner [11]. Security is one of the most important and paramount elements of such a computing environment.

In a cross-clouds federated environment, security concerns are even more important and complex. Cloud computing paradigm, in general, will only be adopted by the users, if they are confident that their data and privacy are secured. Cloud computing involves the sharing or storage by users of their own information on remote servers owned or operated by others and accesses through the Internet or other connections. Cloud computing services exist in many variations, including data storage sites, video sites, tax preparation sites, personal health record websites and many more. The entire contents of a user's storage device may be stored with a single cloud provider or with many cloud providers. Whenever an individual, a business, a government agency, or any other entity shares information in the cloud, privacy or confidentiality questions arise which should be properly addressed to tap the market among various cloud players.

## IV. PROPOSED SCHEME

Our secure data sharing scheme for Federated cloud contains various cloud instances belonging to same Cloud host or different hosts that participate in computation to get overall benefit which is not possible with a single cloud. Each cloud instance will share their data secretly without knowing other hosts data thus ensuring privacy and achieve the final result. Cloud host providers Exchanges data to solve the $n^2$ problem by facilitating as mediators for enabling connectivity among disparate cloud environments.

In our proposed scheme whenever customer requests cloud host provider for service, also if it is an complex application request and the computation depends on other cloud hosts values then it is required to form into federation of clouds as shown in figure-2 above. Among the cloud one will act as Trusted Cloud authority (TCA) which will control and coordinate entire computation. TCA will request will accepts credential / if already contains credentials of each cloud it will use it to initialize the secure data sharing scheme by giving secret keys and initiate the process. The various phases of working in our proposed scheme are described in the next section and outlined diagrammatically in the given figure-3.

Upon request from client/application TCA will creates a Session for that particular instance of computation and session-id's are dynamically created for each host participating in computation. Session-id's are sent to all the cloud hosts in federation privately. Session-id can be used for authentication when each of them exchange data during computation. Internally cloud hosts will have co-coordinators to coordinate the computation which will work according to SLA. Our scheme uses SMC[12] mechanism but the secret value used in data sharing is encrypted which is difficult to know as we have used DL technique and finally each cloud can decrypt the final value by using their secret keys. In our scheme secret value will not be known to the TCA also, as it is encrypted by hosts with their own keys.

① Credentials
② Private Key $g_i$
③ Generation of Secret Primitive Polynomial
④ SMC implementation to compute Sum Polynomial
⑤ Public keys $h_i$, $t_i$ for individual verification and $\delta$ for secret recovery
⑥ Malicious Cloud Verification
⑦ Report Malicious Cloud
⑧ Recover Secret from SUM Polynomial

*Figure 3: Proposed secure data sharing in Federated Clouds*

## V. WORKING OF PROPOSED SCHEME

The proposed scheme is used to secure secret data when shared during computation between federated clouds. In this scheme the secret data is encrypted and decrypted by the each cloud to retrieve original value. We assume that following assumptions hold good at initialization phase.

1. That TCA and cloud hosts providers exchange data securely
2. All Cloud providers are honest without malicious in nature.

The data sharing scheme works in following phases as

1. Initialization Phase
2. Distribution Phase
3. Verification Phase
4. Recovery Phase

*A. Initialization Phase*

In this phase TCA will starts session and session id's are sent to all clouds secretly that participate in computation. Then TCA by using their credentials computes and sends private and public keys for cloud hosts in federation for computation.

Let $C_1, C_2, C_3, \ldots\ldots\ldots\ldots..C_n$ are the clouds involved in computation.

1. The credentials of each cloud $C_i$ are sent to TCA by $C_1, C_2\ldots.C_n$
2. TCA generates large primes $CP_i$ from credentials of each cloud $C_i$.
3. TCA computes $NP_i = 2*CP_i$
4. For each cloud $C_i$, TCA generates a primitive root '$g_i$' from $NP_i$.
5. TCA sends $g_i$ securely which is private to each cloud $C_i$, and $NP_i$ is public to all the clouds.

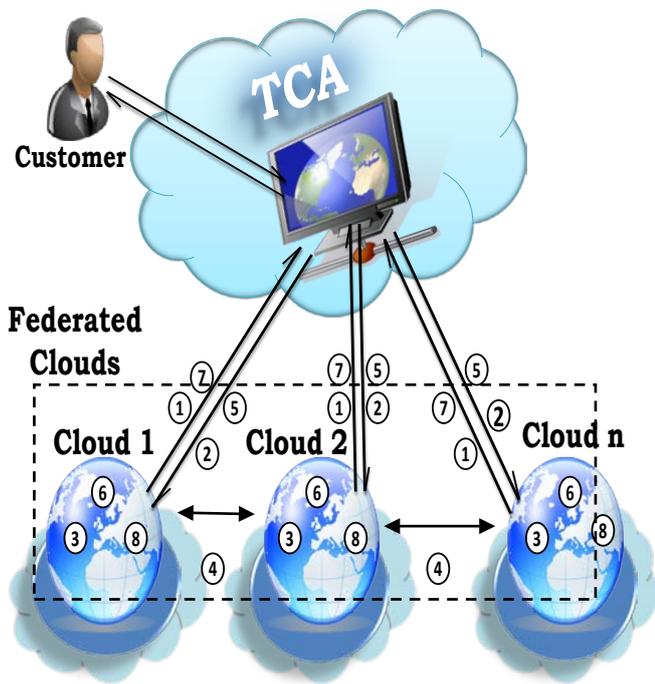

## B. Generation of Polynomial

1. Each cloud $C_i$ generates a group $Z_{Npi}^*$ with the generator $g_i$ and $Np_i$.

2. $C_i$ builds Galois field (GF) consisting of primitive elements with the group $Z_{Npi}^*$ ie., Galois field(ie.,$GF(g_i^{bi})$) has $\Phi(g_i^{bi} - 1)$ primitive elements where $b_i \in Z_{Npi}^*$.

3. Each cloud $C_i$ generates a polynomial $f_i(x)$ with coefficients in GF and hence $f_i(x)$ is a primitive polynomial.

    [ie. $f_i(x) = a_0 x + a_1 x^1 + a_2 x^2 + \ldots + a_{n-1} x^{n-1}$] where $f_i(0) = a_0$

## C. Distribution Phase

In this phase each cloud host in federation exchange secrets for computation to achieve final polynomial with secret value in encrypted form

1. Each Coefficient $a_i$ in primitive polynomial $f_i(x)$ is the primitive number in $GF(g_i^{bi})$ where $0 < i \leq n-1$ and $a_0$ is secret value of $C_i$.

2. Each $C_i$ computes, $a_0 = S_i d_i$ where $d_i = (g_i^{bi})\delta_i$ where $\delta_i \in Z_{Npi}^*$ such that $g_i^{bi} \delta_i \equiv 1 \mod Np_i$ here $S_i$ is the secret that is to be shared between clouds during computation.

3. Each Cloud $C_i$ implements Secure Multiparty Computation (SMC) scheme and computes final sum polynomial $F(x) = \sum_{i=1}^{n} f_i(x)$ and coefficients are in GF sends it to TCA for verification.

## D. Verification Phase

In this phase each cloud host in federation verifies the secret value by decrypting and finds the malicious host if exists and reports to TCA or rejects its value.

**Note:** Any polynomial $f(x)$ with co-efficient of GF(P) satisfies the Identity, $F(x^P) \equiv [f(x)]^P$ (since $g_i = P$ and $GF(P) = GF(g_i)$)

1. TCA randomly selects a prime $gp_i$ that satisfies the identity stated above.

    hence $F(x^{gpi}) \equiv F(x)^{gpi}$

2. Then TCA chooses a small random number $t_i \in Z^+$. $\forall t_i \; \exists h_i \in Z^+ \ni h_i t_i \equiv 1 \pmod{gp_i}$.

3. TCA sends $gp_i$, $h_i$, $t_i$ to the corresponding clouds $C_i$ and announces as public to all the clouds.

4. Each cloud $C_i$ chooses a secret element $r_i \in GF(g_i^{bi})$ such that $X^r_i \equiv h_i (\mod F(x), gp_i)$

5. Each cloud $C_i$ verifies $C_j$ as $X^{r_i t}_{i\,j} \equiv (X^{r}_i)^{t}_j \equiv h_j^{tj} \equiv 1 (\mod (F(x), gp_j))$

6. If any cloud $C_i$ is malicious then the above congruence dissatisfies, since the Sum Polynomial $F(x)$ sent from $C_i$ to $C_j$ is wrong.

    ie. $X^{r_i t}_{i\,j} \neq 1 (\mod F(x), gp_i)$

## E. Recovery Phase

In this phase after verification by each cloud $C_i$, the secret is recovered by using following steps by each party. Secret can be recovered even if there exists a malicious party $m(m<n/2)$.

$S = \sum(S_i d_i)$ where $d_i = (g^{bi})\delta_i$ where $\delta_i \in Z_{npi}^*$ such that $g_i^{bi} \delta_i \equiv 1 \mod np_i$

$S = S_1(g_1^{b_1})\delta_1 + S_2(g_2^{b_2})\delta_2 + \ldots + S_n(g_n^{b_n})\delta_n$.

$= S_1 g_1^{b_1} . \delta_1 + S_2 g_2^{b_2} . \delta_2 + \ldots +$

$+ S_n g_n^{b_n} . \delta_n$

$= S_1(g_1^{b_1} * g_1^{-b_1} \mod np_1) + S_2(g_2^{b_2} * g_2^{-b_2} \mod np_2) + \ldots + S_n(g_n^{b_n} * g_n^{-b_n} \mod np_n)$

$= S_1(g_1^0 \mod np_1) + S_2(g_2^0 \mod np_2) + \ldots + S_n(g_n^0 \mod np_n)$

$= S_1^* 1 + S_2^* 1 + \ldots + S_n^* 1$

$= S_1 + S_2 + \ldots + S_n$

Further in recovery phase SMC can be applied to the following three cases in recovering secret if malicious cloud host exists during data sharing or data recovery when it is distributed among multiple or federated clouds.

*Case 1:* Assume All 'n' clouds hosts in federation are Honest for 'n' honest clouds, The co-efficient of $x_o$ in sum polynomial $F(x)$ is the sum of secret shares of all $C_i$ and it is valid for each $C_i$ iff $X^{r\,t}_{i\,i} \equiv 1(\mod F(x), gp_i)$

*Case 2:* Assume that n-1 cloud hosts in a Federation are Honest with some are malicious

For 'n-1' honest clouds, If any cloud is dishonest among 'n' clouds the 'n-1' clouds together obtains the sum of secret shares as sum of secret shares as

For n-1 parties we reconstruct secret S as

$S_{n-1} = (s_1 g_1^{b_1})^{\delta_1} + (s_2 g_2^{b_2})^{\delta_2} + \ldots + (s_{n-1} g_{n-1}^{b_{n-1}})^{\delta_{n-1}}$.

In the sum Polynomial, the sum of the secrets obtained by each cloud is,

$S = \sum(S_i d_i)$ where $d_i = (g^{bi})\delta_i$ where $\delta_i \in Z_{Npi}^*$ such that $g_i^{bi} \delta_i \equiv 1 \mod Np_i$

$S = S_1(g_1^{b_1})\delta_1 + S_2(g_2^{b_2})\delta_2 + \ldots + S_n(g_n^{b_n})\delta_n$.

$= S_1 g_1^{b_1} . \delta_1 + S_2 g_2^{b_2} . \delta_2 + \ldots$

$+ S_n g_n^{b_n} . \delta_n$

$S = S_{n-1} + S_n g_n^{b_n} . \delta_n$

ie., $S_n g_n^{b_n} . \delta_n = S - S_{n-1}$

*Case 3:* Assuming that there are >=n/2 cloud hosts are malicious in federation.

If n/2 are malicious clouds then

$S = S_1(g_1^{b_1})\delta_1 + S_2(g_2^{b_2})\delta_2 + \ldots\ldots +$
$S_2(g_{n/2}^{b_{n/2}})\delta_{n/2} + \ldots\ldots + S_n(g_n^{b_n})\delta_n$
$S = S_1(g_1^{b_1})\delta_1 + S_2(g_2^{b_2})\delta_2 + \ldots\ldots +$
$S_2(g_{n/2}^{b_{n/2}})\delta_{n/2} + \ldots\ldots + S_n(g_n^{b_n})\delta_n + S_n(g_n^{b_n})\delta_n$
$S = 4*(n/2) \text{ unknowns} + \ldots\ldots\ldots\ldots + S_{n-1} + S_n$
$\therefore S \neq S_{n/2}$

The unknowns in the sum polynomial are 2n, so it is not possible to get S from 2n unknowns.

## VI. EXPERIMENTAL ANALYSIS OF PROPOSED SCHEME

We have verified the only the base scheme used in data sharing between the clouds by using Java 1.7 on Intel Core-i3 processor with 4 GB RAM. We have taken only small values as credentials due to computation resource constraint which has given following results, here number of clouds in federation is taken as 4.

Enter how many Clouds involve in Federation for Communication:     4

### A. Generation of Parameters:

*Enter the grant type:* Client
*Enter the service type:* Application
*Enter the client name:* Amazon
*Enter the client region:* Asia
*Enter the client location:* India
*Enter the service payment:* 250000000
*Enter the service expiry date:* 31-Dec-2025

Cp = 4327    Np1 = 8654  g1 = 8647

*Enter the grant type:* Client
*Enter the service type:* Application
*Enter the client name:* Google Docs
*Enter the client region:* America
*Enter the client location:* Mexico City
*Enter the service payment:* 3000000000
*Enter the service expiry date:* 31-Dec-2030

Cp = 5669    Np2 = 11338    g2 = 11311

*Enter the grant type:*  Client
*Enter the service type:* Application
*Enter the client name:* Google Cloud Services
*Enter the client region:* Asia
*Enter the client location:* Pakistan
*Enter the service payment:* 300000000000
*Enter the service expiry date:* 31-Dec-2025

Cp = 6203    Np3 = 12406    g3 = 12401

*Enter the grant type:* Client
*Enter the service type:* Application
*Enter the client name:* HP Cloud Provider
*Enter the client region:* Asia

*Enter the client location:* Bangladesh
*Enter the service payment:* 3600000000
*Enter the service expiry date:* 31-Dec-2035

Cp = 5843    Np4 = 11686    g4 = 11681

### B. Generation of Polynomials:

$(7)X^3 + (26)X^2 + (6)X^1 + (2)X^0$
$(19)X^3 + (16)X^2 + (12)X^1 + (4)X^0$
$(10)X^3 + (13)X^2 + (3)X^1 + (6)X^0$
$(24)X^3 + (15)X^2 + (19)X^1 + (8)X^0$

### C. Distribution of Secret:

s1=2 (original secret)
s2=4 (original secret)
a0= $s_1d_1$= 646541456023        (E)encrypted)
a0= $s_2d_2$= 1636831633111541     (E)encrypted)
s3=6 (original secret)
s4=8 (original secret)
a0= $s_3d_3$= 2932807359995777662001(E)
a0= $s_4d_4$= 2540271545712591010246081(E)

where $d_i=(g_i^{bi})\delta_i$ where $\delta_i \in Z_{Np_i}^*$ such that $g_i^{bi}\delta_i \equiv 1 \mod Np_i \Rightarrow \delta_i = g_i^{-bi} \mod Np_i$

The **revised** polynomials are:

$(24)X^3 + (4)X^2 + (20)X^1 + (8368306130700080)X^0$
$(3)X^3 + (18)X^2 + (23)X^1 + (2076343186244444682973568)X^0$
$(18)X^3 + (24)X^2 + (20)X^1 + (21783804456699014989946336906386176)X^0$
$(11)X^3 + (4)X^2 + (24)X^1 + (16408063398992467575067769015170019871641600)X^0$

The Sum of the Polynomials obtained at each party is

$(56)X^3 + (50)X^2 + (87)X^1 +$ **16408063420776272031766784005116356778027776** $)X^0$ (encrypted value) **original values is (20)**

### D. Recovery of Secret:

*Case 1:* Assuming there are no malicious cloud host in Federation of clouds

$S = \sum(S_id_i) \; i=1,2,3,4$
$S = s_1d_1 + s_2d_2 + s_3d_3 + s_4d_4$
$S = S_1(g_1^{b_1})\delta_1 + S_2(g_2^{b_2})\delta_2 + S_3(g_3^{b_3})\delta_3 + S_4(g_4^{b_4})\delta_4$
$= S_1 g_1^{b_1}.\delta_1 + S_2 g_2^{b_2}.\delta_2 + S_3 g_3^{b_3}.\delta_3 + S_4 g_4^{b_4}.\delta_4$
$= S_1(g_1^{b_1} * g_1^{-b_1} \mod np_1) + S_2(g_2^{b_2} * g_2^{-b_2} \mod np_2) + S_3(g_3^{b_3} * g_3^{-b_3} \mod np_3) + S_4(g_4^{b_4} * g_4^{-b_4} \mod np_4)$
$= S_1(g_1^0 \mod np_1) + S_2(g_2^0 \mod np_2) + S_2(g_3^0 \mod np_3) + S_4(g_4^0 \mod np_4)$
$= S_1*1 + S_2*1 + S_3*1 + S_4*1$
$= S_1 + S_2 + S_3 + S_4$
$S = 2+4+6+8$
$S = 20$

*Case 2:* Assuming honest clouds in federation are $\leq n-1$

$S_0 = \sum(S_i d_i)$ $i=1,2,3$

$S_0 = s_1 d_1 + s_2 d_2 + s_3 d_3$

$S_0 = S_1(g_1^{b_1})\delta_1 + S_2(g_2^{b_2})\delta_2 + S_3(g_3^{b_3})\delta_3$

$\phantom{S_0} = S_1 g_1^{b_1} \cdot \delta_1 + S_2 g_2^{b_2} \cdot \delta_2 + S_3 g_3^{b_3} \cdot \delta_3$

$\phantom{S_0} = S_1(g_1^{b_1} * g_1^{-b_1} \bmod np_1) + S_n(g_2^{b_2} * g_2^{-b_2} \bmod np_2) + S_3(g_3^{b_3} * g_3^{-b_3} \bmod np_3)$

$\phantom{S_0} = S_1(g_1^0 \bmod np_1) + S_2(g_2^0 \bmod np_2) + S_2(g_3^0 \bmod np_3)$

$\phantom{S_0} = S_1 * 1 + S_2 * 1 + S_3 * 1$

$\phantom{S_0} = S_1 + S_2 + S_3$

$S_0 = 2+4+6$

$S_0 = 12$

The original Sum of Secrets is, $S=20$

$\quad S = S_0 + S_4^{d4}$
$\quad 20 = 12 + S_4^{d4}$
$\quad S_4^{d4} = 20-12c$
$\quad S_4^{d4} = 8$

Therefore, $\quad S = S_0 + S_4^{d4}$
$\quad\quad\quad S = 12+8$
$\quad\quad\quad S = 20$

The Sum of the Polynomials after recovering the secret at each party is ::

$(56)X^3 + (50)X^2 + (87)X^1 + (20)X^0$

*Case 3:* Assuming we are having $n/2$ or $(n-1)/2$ are malicious clouds

$S = S_1(g_1^{b_1})\delta_1 + S_2(g_2^{b_2})\delta_2 + \ldots + S_2(g_{n/2}^{b_{n/2}})\delta_{n/2} + \ldots + S_n(g_n^{b_n})\delta_n$

$S = S_1(g_1^{b_1})\delta_1 + S_2(g_2^{b_2})\delta_2 + \ldots + S_2(g_{n/2}^{b_{n/2}})\delta_{n/2} + \ldots + S_n(g_n^{b_n})\delta_n + S_n(g_n^{b_n})\delta_n$

$S = 4*(n/2)$ unknowns$+ \ldots + S_{n-1} + S_n$

$\therefore S \neq S_{n/2}$

The unknowns in the sum polynomial are $2n$, so it is not possible to get S from $2n$ unknowns.

## VII. USE CASES

In Weather Research and Forecasting application used for Agriculture or for any governmental purposes uses values from different cloud host stations at different locations to analyses the final result which works in federation. Here data should be correct and secure so that it may not give wrong results which may lead to disaster.

For forecasting stations, due to the nature of certain weather phenomena such as hurricanes or tornadoes, performing accurate predictions in very short time spans is vital to make appropriate preparations involving business operations management and government and human related logistics. Thus, sharing of resources between institutions to provide elasticity and dynamic capacity in extreme situations is key.

The applications like Online Voting or Online Bidding or Real time Game playing stations when deployed on clouds uses multiple hosts at located at different geographical areas will demands data to have privacy and secure.

## VIII. CONCLUSION

Cloud computing key role in IT sector in delivering services at low cost and in an effective manner. Clouds should form into federation in order to perform computation collectively to achieve a result. At the same time the security threats like data should be addressed with by using novel techniques. In this paper we have used threshold data sharing technique to be used in federation of clouds which allows data privacy and security in transit between them. We have analyzed the base scheme and results are noted. The same technique can be used to recover data when distributed between multiple clouds and one of the cloud host was not available due to natural disaster or technical problem thus provides solution to data availability in cloud computing. In future we try to implement this technique on real time cloud and also for authenticating automated applications running on clouds.